\documentclass[a4paper]{article}

\usepackage{INTERSPEECH2022}

\usepackage{amssymb}
\usepackage{enumerate}
\usepackage{comment}
\usepackage{color}
\usepackage{bm}
\usepackage{array}
\usepackage{multirow}
\usepackage{colortbl}
\usepackage{url}
\usepackage[subrefformat=parens]{subcaption}

\newlength\savedwidth
\newcommand{\wcline}[1]{\noalign{\global\savedwidth\arrayrulewidth\global\arrayrulewidth 1.0pt} \cline{#1}
\noalign{\global\arrayrulewidth\savedwidth}}

\title{How Information on Acoustic Scenes and Sound Events\\Mutually Benefits Event Detection and Scene Classification Tasks}
\name{Keisuke Imoto$^{1}$, Yuka Komatsu$^{1}$, Shunsuke Tsubaki$^{1}$, and Tatsuya Komatsu$^{2}$}
\address{
$^1$Doshisha University, Japan\\
$^2$LINE Corporation, Japan. 
}
\email{keisuke.imoto@ieee.org}

\begin{document}

\maketitle
%
\begin{abstract}
Acoustic scene classification (ASC) and sound event detection (SED) are fundamental tasks in environmental sound analysis, and many methods based on deep learning have been proposed. Considering that information on acoustic scenes and sound events helps SED and ASC mutually, some researchers have proposed a joint analysis of acoustic scenes and sound events by multitask learning (MTL). However, conventional works have not investigated in detail how acoustic scenes and sound events mutually benefit SED and ASC. We, therefore, investigate the impact of information on acoustic scenes and sound events on the performance of SED and ASC by using domain adversarial training based on a gradient reversal layer (GRL) or model training with fake labels. Experimental results obtained using the TUT Acoustic Scenes 2016/2017 and TUT Sound Events 2016/2017 show that pieces of information on acoustic scenes and sound events are effectively used to detect sound events and classify acoustic scenes, respectively. Moreover, upon comparing GRL- and fake-label-based methods with single-task-based ASC and SED methods, single-task-based methods are found to achieve better performance. This result implies that even when using single-task-based ASC and SED methods, information on acoustic scenes may be implicitly utilized for SED and vice versa.
\end{abstract}
\noindent\textbf{Index Terms}: acoustic scene analysis, sound event detection, multitask learning, domain adversarial training, gradient reversal layer, fake label
%
%
\section{Introduction}
\label{sec:intro}
Environmental sound analysis \cite{Virtanen_Springer2018_01,Imoto_AST2018_01} is the analysis of audio recordings that are not limited to voice or music and has various real-world applications such as machine condition monitoring, automatic surveillance, media tagging, and biomonitoring systems \cite{Koizumi_DCASE2020_01,Chan_EUSIPCO2010_01,Fonseca_DCASE2018_01,Morfi_JASA2021_01}.
In environmental sound analysis, acoustic scene classification (ASC) and sound event detection (SED) are the fundamental research topics.
In ASC, an acoustic scene label is estimated from a sound recording where the acoustic scene is defined as the place or situation in which the audio is recorded, such as \textit{home}, \textit{train}, or \textit{meeting}.
In SED, sound event labels and their timestamps in the sound recording are predicted, where a sound event is defined as a sound class, such as \textit{bird singing}, \textit{cutlery}, or \textit{car horn}.

Recently, many systems for environmental sound analysis have been implemented using neural networks.
For example, Valenti et al. \cite{Valenti_IJCNN2017_01}, Liping et al. \cite{Liping_DCASE2018_01}, Tanabe et al. \cite{Tanabe_DCASE2018_01}, and Raveh and Amar \cite{Raveh_DCASE2018_01} introduced ASC methods based on the convolutional neural network (CNN), Xception, VGG, and ResNet, respectively.
\c{C}ak\i r et al. introduced the convolutional recurrent neural network (CRNN), which can capture temporal and spectral information of sound events \cite{Cakir_TASLP2017_01}, for the SED task.
Kong et al. \cite{Kong_TASLP2020_01} and Miyazaki et al. \cite{Miyazaki_DCASE2020_01} proposed SED methods using a Transformer and Conformer encoder, respectively.

\begin{figure*}[t]
\centering
\vspace{7pt}
\includegraphics[width=1.94\columnwidth]{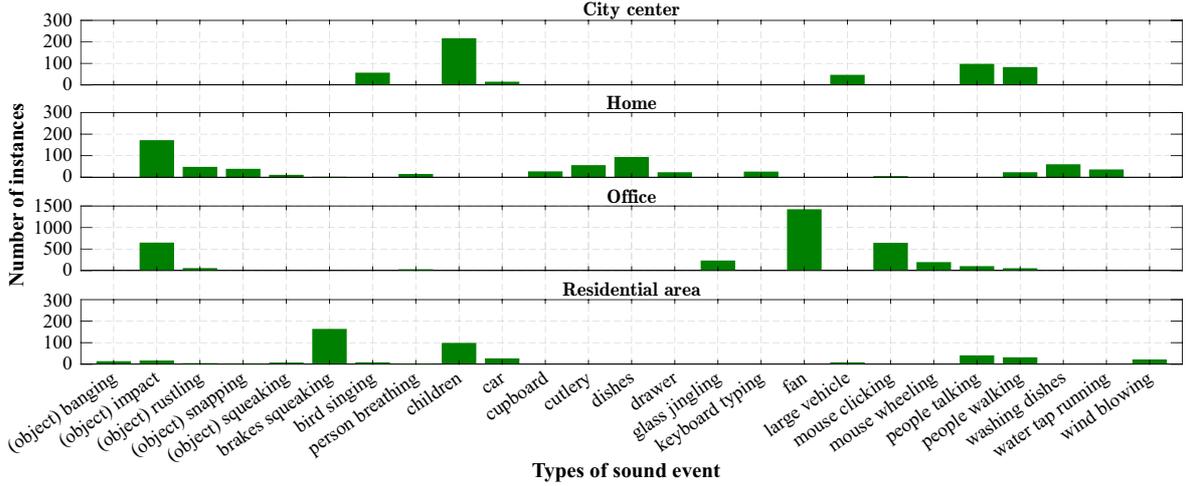}
\vspace{0pt}
\caption{Frequency of sound event instances for each acoustic scene in dataset used for evaluation experiments}
\label{fig:numofinstance}
\vspace{14pt}
\end{figure*}

The conventional methods for environmental sound analysis address scene classification and event detection tasks separately.
However, acoustic scenes and sound events are closely related; for example, in the acoustic scene \textit{office}, the sound events \textit{keyboard typing} and \textit{mouse clicking} tend to occur, whereas the sound events \textit{cutlery} and \textit{car horn} occur infrequently, as shown in Fig.~\ref{fig:numofinstance}.
Thus, the information on the sound events \textit{keyboard typing} and \textit{mouse clicking} will help in estimating the acoustic scene \textit{office} and vice versa.
Considering the relationship between acoustic scenes and sound events, Mesaros et al. \cite{Mesaros_EUSIPCO2011_01} and Heittola et al. \cite{Heittola_JASM2013_01} proposed SED methods utilizing information on acoustic scenes.
Imoto and co-workers proposed ASC methods based on Bayesian generative models, in which information on sound events is considered \cite{Imoto_IEICE2016_01,Imoto_TASLP2019_01}.
Bear et al. \cite{Bear_INTERSPEECH2019_01}, Tonami et al. \cite{Tonami_IEICE2021_01}, and Jung et al. \cite{Jung_ICASSP2021_01} presented methods of jointly analyzing acoustic scenes and sound events based on the multitask learning (MTL) of ASC and SED.
These works have revealed that utilizing the relationship between acoustic scenes and sound events improves the performance of each downstream task.
However, conventional studies have not fully clarified how acoustic scenes and sound events mutually benefit event detection and scene classification tasks.
In this work, we thus investigate the impact of information on acoustic scenes and sound events on the performance of SED and ASC by using domain adversarial training with a gradient reversal layer (GRL) \cite{Ganin_JMLR2016_01} or model training with fake labels.

The remainder of this paper is structured as follows.
In section 2, we overview the conventional scene classification and event detection methods based on single and multitask learning.
Section 3 introduces the model training based on GRL and fake labels.
In section 4, we discuss in detail how information on acoustic scenes and events affects the performance of ASC and SED, referring to the results of evaluation experiments.
We conclude this paper in section 5.
%
%
%
%
%
%
%
\section{Conventional Methods}
\label{sec:conventional}
\subsection{Conventional Scene Classification and Event Detection Methods}
\label{ssec:ConvASCandSED}
In this section, we overview the conventional works of ASC and SED.
Recently, many neural-network-based approaches such as CNN \cite{Valenti_IJCNN2017_01,Hershey_ICASSP2017_01}, CRNN \cite{Cakir_TASLP2017_01}, and Transformer encoder \cite{Kong_TASLP2020_01} have been proposed.
These methods use the time--frequency representation of the acoustic signal $X \in \mathcal{R}^{D \times T}$, such as the log mel-band spectrogram, as the acoustic feature, where $D$ and $T$ indicate the numbers of frequency bins and time frames, respectively.
This acoustic feature is then fed to the ASC or SED network.

The model parameters of ASC are trained using the network output $y_{n}$ and the cross-entropy (CE) loss function ${\mathcal L}_{{\rm scene}}$,

%
\begin{align}
{\mathcal L}_{{\rm scene}} &= - \! \sum^{N}_{n=1} {\Big \{} s_{n} \log ( y_{n} ) {\Big \}},
\label{eq:scene_loss}
\end{align}
%

\noindent where $N$ and $s_{n}$ are the number of acoustic scene classes and the target scene label, respectively.

The model parameters of SED are trained using the network output $y_{t,m}$ and the following binary cross-entropy (BCE) loss function ${\mathcal L}_{{\rm event}}$,

%
\begin{align}
\hspace{0pt} {\mathcal L}_{{\rm event}} &= - \! \sum^{T}_{t=1} \hspace{-1pt} {\Big \{} {\bf z}_{t} \log ( {\bf y}_{t} ) \! + \! (1 \! - \! {\bf z}_{t}) \log (1 \! - \! {\bf y}_{t} ) \hspace{-1pt} {\Big \}} \nonumber\\[2pt]
&\hspace{0pt} = - \!\!\! \sum^{T\!, \hspace{1pt} M}_{t,m=1} \hspace{-4pt} {\Big \{} \hspace{-0.5pt} z_{t,m} \log ( y_{t,m} ) \! + \! (1 \! - \! z_{t,m}) \log (1 \! - \! y_{t,m}) \hspace{-1.2pt} {\Big \}} \hspace{-0.3pt} ,\hspace{-3pt}
\label{eq:event_loss}
\end{align}
%

\noindent where $T$, $M$, and $z_{t,m}$ indicate the number of time frames, the number of sound event classes, and the target event label, respectively.
%
%
%
\begin{figure}[t!]
\centering
\vspace{3pt}
\centering
\includegraphics[width=0.93\columnwidth]{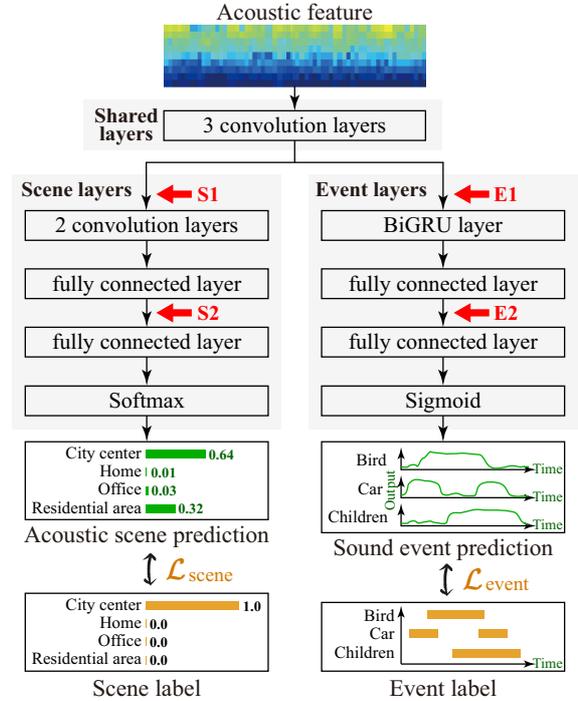}
\vspace{3pt}
\caption{Network structure of MTL-based joint analysis of ASC and SED \cite{Tonami_IEICE2021_01}}
\label{fig:conventionalMTL}
\vspace{10pt}
\end{figure}
%
%
%
\subsection{Joint Analysis of Acoustic Scenes and Sound Events Based on Multitask Learning}
\label{ssec:ConvMTL}
Most conventional works address ASC and SED separately; however, many acoustic scenes and sound events are related mutually.
Considering that knowledge of acoustic scenes and sound events can mutually aid in their estimation, the joint analysis of acoustic scenes and sound events based on multitask learning has been proposed \cite{Bear_INTERSPEECH2019_01,Tonami_IEICE2021_01,Imoto_ICASSP2020_01}.

The network structure of MTL for the joint analysis of acoustic scenes and sound events is shown in Fig.~\ref{fig:conventionalMTL}.
The MTL network consists of shared, scene-specific, and event-specific layers.
The shared layers have convolution layers for extracting the embedded acoustic features and are expected to extract the information common to acoustic scenes and sound events.
The scene- and event-specific layers include the CNN and recurrent neural network (RNN) layers, which are used for the downstream tasks of ASC and SED.

To train the MTL model of ASC and SED, the following loss function ${\mathcal L}$ is applied:

%
\begin{align}
{\mathcal L} &= \alpha {\mathcal L}_{{\rm scene}} + \beta {\mathcal L}_{{\rm event}},
\label{eq:mtl_loss}
\end{align}
%
\ \\[-10pt]

\noindent where $\alpha$ and $\beta$ are the constant weighting factors of ASC and SED losses, respectively.
In this work, $\beta = 1.0$ can be set without loss of generality.
%
%
%
\section{Methodology}
\label{sec:methodology}
Some conventional studies on jointly analyzing acoustic scenes and sound events revealed that pieces of information on acoustic scenes and sound events improve the performance of SED and ASC, respectively \cite{Tonami_IEICE2021_01}.
However, how acoustic scenes and sound events benefit SED and ASC tasks mutually has not been fully investigated in conventional studies.
To evaluate how information on acoustic scenes and sound events benefits the performance of ASC and SED, we apply the domain adversarial training based on GLR and fake labels of acoustic scenes and sound events, which enables the training of the model without intentionally using information on acoustic scenes and sound events.
%
\subsection{MTL Based on Domain Adversarial Training}
\label{ssec:DAT}
We first apply the domain adversarial training based on GLR \cite{Ganin_JMLR2016_01} to the MTL of ASC and SED.
The GRL acts as an identity transformation of the input during the forward propagation, but changes the sign of input, i.e., it multiplies by -1 during the backpropagation as follows.

\vspace{-5pt}
\begin{align}
&\ \hspace{-23pt} \textrm{Forward:} \ \ \ \ \ \ \hspace{-1.5pt} G(\textbf{x}) = \textbf{x}\\[2pt]
&\ \hspace{-29pt}\textrm{Backward:} \ \ \ \ \ \ \hspace{3.2pt} \frac{dG}{dx} = - \lambda \textbf{I}
\end{align}
\vspace{0pt}

\noindent Here $\textbf{x}$, $\lambda$, and $\textbf{I}$ are the input of the GRL, the weighting factor, and the identity matrix, respectively.
By adding the GRL to a network, the model parameters in layers prior to the GRL are trained as increasing the prediction error.
Thus, the GRL enables the training of models that do not depend on the downstream task with this layer.
In the experiments conducted in this work, we added the GRL to either position shown by the red arrows (S1, S2, E1, and E2) in Fig.~\ref{fig:conventionalMTL} and evaluated the performance of ASC and SED.
%
%
\subsection{MTL Based on Fake Label}
\label{ssec:fake}
We also evaluate the performance of ASC and SED using the MTL model trained with fake labels.
In this work, we created fake scene and event labels as follows.

\vspace{-2pt}
\begin{align}
&\ \hspace{-11pt} \textrm{Fake scene label:} \ \ \ \ \ \hspace{6.4pt} \hat{\textbf{s}} = \textrm{shuffle} (\textbf{s})\\[2pt]
&\ \hspace{-10pt} \textrm{Fake event label:} \ \ \ \ \ \ \hat{\textbf{z}}_{t} = \textrm{shuffle} (\textbf{z}_{t})
\end{align}
\vspace{-1pt}

\noindent Here, shuffle, $\textbf{s}$, and $\textbf{z}_{t}$ indicate the random reordering operation of vector elements, the target scene label vector, and the target event label vector in time frame $t$, respectively.
In the experiment, we replaced either the acoustic scene or sound event labels with the fake labels.
As with the GRL, the model training with fake labels allows us to investigate how acoustic scenes and sound events affect ASC and SED.
%
%
%
\vspace{5pt}
\section{Evaluation Experiments}
\label{sec:experiment}
\subsection{Experimental Conditions}
\label{ssec:condition}
We conducted evaluation experiments to investigate how the information on acoustic scenes and sound events can benefit ASC and SED tasks.
In the evaluation experiments, we used a dataset composed of parts of the TUT Acoustic Scenes 2016/2017 and TUT Sound Events 2016/2017 \cite{Mesaros_EUSIPCO2016_01,Mesaros_DCASE2017_01}.
From these four datasets, we selected sound clips of four acoustic scenes, \textit{city center}, \textit{home}, \textit{office}, and \textit{residential area}, which include 25 sound event classes.
The data comprises a total of 266 min of sounds (192 min for the development set and 74 min for the evaluation set).
The details of the dataset used for the evaluation experiments are found in \cite{Imoto_dataset2019_01}.

As an acoustic feature, we extracted the 64-dimensional log mel-band energy, which was calculated with a 40 ms frame length and a 20 ms hop size.
We used the same MTL network structure as that of the conventional method \cite{Tonami_IEICE2021_01}, as shown in Table~\ref{tbl:networks}, where BiGRU and FC layers indicate the bidirectional gated recurrent unit and fully connected layers, respectively.
For each method, we conducted the evaluation experiments 10 times with random initial values of model parameters.
Other experimental conditions are listed in Table~\ref{tbl:parameter}.
\begin{table}[t]
\vspace{5pt}
\footnotesize
\caption{Details of structure of MTL network of ASC and SED}
\vspace{-15pt}
\label{tbl:networks}
\begin{center}
\begin{tabular}{ccc}
\wcline{1-3}
\!\!&\!\!\\[-7pt]
\multicolumn{3}{c}{\textbf{Shared layers}}\\
\wcline{1-3}
\!\!&\!\!\\[-7pt]
\multicolumn{3}{c}{Log-mel energy}\\[1pt]
\multicolumn{3}{c}{500 frames $\times$ 64 mel bin}\\[1pt]
\cline{1-3}
\!&\\[-7pt]
\multicolumn{3}{c}{3$\times$3 kernel size / 128 ch.}\\[1pt]
\multicolumn{3}{c}{Batch norm., Leaky ReLU}\\[1pt]
\multicolumn{3}{c}{1$\times$8 Max pooling}\\
\cline{1-3}
\!&\\[-7pt]
\multicolumn{3}{c}{$\begin{pmatrix} \textrm{3$\times$3 kernel size / 128 ch.}\\[1pt]
\textrm{Batch norm., Leaky ReLU}\\[-1pt]
\textrm{1$\times$2 Max pooling}
\end{pmatrix}$ $\times$ 2
}\\[0pt]
\!&\\[-7pt]
\wcline{1-3}
\multicolumn{1}{c}{}\\[-7pt]
\multicolumn{1}{c}{\textbf{Scene layers}}&\!\!&\textbf{Event layers}\\
\wcline{1-3}
\multicolumn{1}{c}{}\\[-7pt]
\multicolumn{1}{c}{3$\times$3 kernel size / 256 ch.}&\!\!&\\[1pt]
\multicolumn{1}{c}{Batch norm., Leaky ReLU}&\!\!&BiGRU w/ 32 units\\[1pt]
\multicolumn{1}{c}{25$\times$1 Max pooling}&\!\!&\\
\cline{1-1} \cline{3-3}
\multicolumn{1}{c}{}\\[-7pt]
\multicolumn{1}{c}{3$\times$3 kernel size / 256 ch.}&\!\!&\multirow{2}{*}{}\\[1pt]
\multicolumn{1}{c}{Batch norm., Leaky ReLU}&\!\!&FC w/ 32 units, Leaky ReLU\\[1pt]
\multicolumn{1}{c}{Global max pooling}&\!\!&\\
\cline{1-1} \cline{3-3}
\multicolumn{1}{c}{}\\[-7pt]
\multicolumn{1}{c}{FC w/ 32 units, Leaky ReLU}&\!\!&FC w/ 25 units, Sigmoid\\[0pt]
\multicolumn{1}{c}{}\\[-7.5pt]
\cline{1-1}\wcline{3-3}
\multicolumn{1}{c}{}\\[-7pt]
\multicolumn{1}{c}{FC w/ 4 units, Softmax}&\!\!&\\
\wcline{1-1}
\end{tabular}
\end{center}
\end{table}
\begin{table}[t]
\vspace{0pt}
\footnotesize
\caption{Experimental conditions}
\vspace{-15pt}
\label{tbl:parameter}
\begin{center}
\begin{tabular}{ll}
\wcline{1-2}
&\\[-7pt]
Acoustic feature & Log-mel energy (64 dim.)\\
Frame length \hspace{-3pt} / \hspace{-3pt} shift & 40 ms \hspace{-3pt} / \hspace{-3pt} 20 ms\\
Length of sound clip & 10 s\\
Optimizer & RAdam \cite{Liu_ICLR2020_01}\\[0pt]
$\alpha$, $\beta$ & 0.0001, 1.0\\[0pt]
$\lambda$ & 1.0\\[0pt]
\wcline{1-2}
\end{tabular}
\vspace{6pt}
\end{center}
\end{table}
%
%
%
\subsection{Experimental Results}
\label{ssec:result}
\subsubsection{Overall Performance of ASC and SED}
\label{sssec:result1}
Table~\ref{tbl:performance01} shows the overall performance of ASC and SED.
CNN (ASC) and CNN-BiGRU (SED) are the single task networks for ASC and SED; they have the same network structures as the shared + scene layers and shared + event layers in Table~\ref{tbl:networks}, respectively.
GRL (S1) indicates the MTL-based method with GRL in S1 of Fig.~\ref{fig:conventionalMTL}.
In this experiment, we evaluated the SED system with the frame-based metric because the SED system outputs the frame-wise predictions, and we can understand the basic trends of the system output with this metric.

The results show that when we apply the GRLs in the scene and event layers or train models with fake labels, the performance of ASC and SED tends to decrease.
These results indicate that pieces of information on acoustic scenes and sound events are effectively used to detect sound events and classify acoustic scenes, respectively.
A comparison of GRL- and fake-label-based methods with single-task-based ASC and SED methods reveals that single-task-based methods achieve better performance.
This result implies that even when using single-task-based ASC and SED methods, information on acoustic scenes may be implicitly utilized for SED and vice versa.
\begin{table}[t]
\vspace{5pt}
\footnotesize
\caption{Overall performance of ASC and SED}
\vspace{-5pt}
\label{tbl:performance01}
\centering
\hspace*{-3pt}
\begin{tabular}{lcccrr}
\wcline{1-6}
&\\[-7pt]
&\multicolumn{2}{c}{\textbf{Scene}}&\!\!\!&\multicolumn{2}{c}{\textbf{Event}} \!\!\\
\cline{2-3}\cline{5-6}
&\\[-7pt]
\multicolumn{1}{c}{\textbf{Method}}&\!\!\!\! \textbf{Micro-} \!\!\!\!&\!\!\!\! \textbf{Macro-} \!\!\!\!&\!\!\!& \multicolumn{1}{c}{\textbf{Micro-}} \!\!\!\!& \multicolumn{1}{c}{\textbf{Macro-}} \!\!\\
\multicolumn{1}{c}{} \!\!\!\!& \textbf{Fscore} \!\!\!\!& \textbf{Fscore} \hspace{-10pt}&\!\!\!\!& \multicolumn{1}{c}{\textbf{Fscore}} \!\!\!\!& \multicolumn{1}{c}{\textbf{Fscore}} \!\!\\
\wcline{1-6}
&\\[-7pt]
CNN {\small (ASC)} \hspace{-4pt}\!\!&\!\! 85.00\% \!\!\!&\!\!\! 84.29\% \!\!\!&\!\!\!&\!\!\! - \!\!\!&\!\! - \!\!\\[1pt]
CNN-BiGRU (SED) \hspace{-4pt}&\!\!\! - \!\!\!&\!\!\! - \!\!\!&\!\!\!&\!\!\! 42.54\% \!\!\!&\!\! 11.09\% \!\!\\
Conventional MTL \hspace{-2pt}\!\!&\!\! \textbf{88.57\%} \!\!\!&\!\!\! \textbf{88.85\%} \!\!\!&\!\!\!&\!\!\! \textbf{44.63\%} \!\!\!&\!\! \textbf{11.57\%} \!\!\\
\cline{1-6}
&\\[-7.5pt]
GRL (S1)\hspace{-2pt}\!\!&\!\! 77.89\% \!\!\!&\!\!\! 76.14\% \!\!\!&\!\!\!&\!\!\! 40.08\% \!\!\!&\!\! 9.64\% \!\!\\
GRL (S2)\hspace{-2pt}\!\!&\!\! 33.34\% \!\!\!&\!\!\! 22.03\% \!\!\!&\!\!\!&\!\!\! 39.59\% \!\!\!&\!\! 9.92\% \!\!\\
Fake scene labels\hspace{-2pt}\!\!&\!\! 24.45\% \!\!\!&\!\!\! 11.55\% \!\!\!&\!\!\!&\!\!\! 41.57\% \!\!\!&\!\! 10.43\% \!\!\\
\cline{1-6}
&\\[-7.5pt]
GRL (E1)\hspace{-2pt}\!\!&\!\! 60.90\% \!\!\!&\!\!\! 56.30\% \!\!\!&\!\!\!&\!\!\! 13.07\% \!\!\!&\!\! 1.12\% \!\!\\
GRL (E2)\hspace{-2pt}\!\!&\!\! 75.23\% \!\!\!&\!\!\! 73.99\% \!\!\!&\!\!\!&\!\!\! 0.02\% \!\!\!&\!\! 0.02\% \!\!\\
Fake event labels\hspace{-2pt}\!\!&\!\! 84.22\% \!\!\!&\!\!\! 84.09\% \!\!\!&\!\!\!&\!\!\! 0.00\% \!\!\!&\!\! 0.00\% \!\!\\
\wcline{1-6}
\end{tabular}
\vspace{5pt}
\end{table}
\begin{table}[t!]
\vspace{0pt}
\footnotesize
\caption{Average Fscores for selected sound events}
\vspace{-5pt}
\label{tab:eventresult1}
\hspace*{-3pt}
\centering
\begin{tabular}{lccccr}
\wcline{1-6}\\[-7pt]
\!\!&\!\!{\bf  bird}\!\!&\!\! \!\!&\!\! \!\!\!&\!\!\!{\bf  keyboard}\!\!\!&\multicolumn{1}{c}{\!\!\!\textbf{large}}\!\!\\
\multicolumn{1}{c}{\multirow{-1.9}{*}{\!\! \bf Method}}&\!\!{\bf singing}\!\!&\!\!\!\multirow{-1.9}{*}{\!\bf  car}\!\!\!&\!\!\!\multirow{-1.9}{*}{\!\bf dishes}\!\!\!&\!\!\!{\bf typing}\!\!\!\!&\multicolumn{1}{c}{\!\!\!\!{\bf vehicle}}\!\!\\
\ \\[-8.5pt]
\wcline{1-6}\\[-7pt]
\ \\[-8.5pt]
\!\!CNN-BiGRU {\scriptsize (SED)}\!\!&\!\!\textbf{46.36\%}\!\!&\!\!\!44.53\%\!\!\!&\!\!0.17\%\!\!\!&\!\!\!\!4.34\%\!\!\!&\!\!\!12.27\%\!\!\\
\ \\[-8.5pt]
\!\!Conv. MTL\!\!&\!\!46.29\%\!\!&\!\!\!\textbf{45.51\%}\!\!\!&\!\!\textbf{0.25\%}\!\!\!&\!\!\!\!\textbf{5.08\%}\!\!\!\!&\!\!\!\!\textbf{12.29\%}\!\!\\
\ \\[-8.5pt]
\!\!GRL (S1)\!\!&\!\!40.96\%\!\!&\!\!\!43.46\%\!\!\!&\!\!0.00\%\!\!\!&\!\!\!\!0.56\%\!\!\!\!&\!\!\!\!9.28\%\!\!\\
\ \\[-8.5pt]
\!\!GRL (S2)\!\!&\!\!37.62\%\!\!&\!\!\!40.12\%\!\!\!&\!\!0.01\%\!\!\!&\!\!\!\!1.21\%\!\!\!\!&\!\!\!\!10.54\%\!\!\\
\ \\[-8.5pt]
\!\!Fake scene label\!\!&\!\!28.93\%\!\!&\!\!\!39.96\%\!\!\!&\!\!0.06\%\!\!\!&\!\!\!\!1.63\%\!\!\!\!&\!\!\!\!12.05\%\!\!\\
\wcline{1-6}
\end{tabular}
\vspace{10pt}
\end{table}
%
%
%
%
\begin{figure}[t]
\vspace{5pt}
\begin{minipage}[b]{0.48\linewidth}
\hspace*{-9pt}
\centering
\includegraphics[width=1.11\columnwidth]{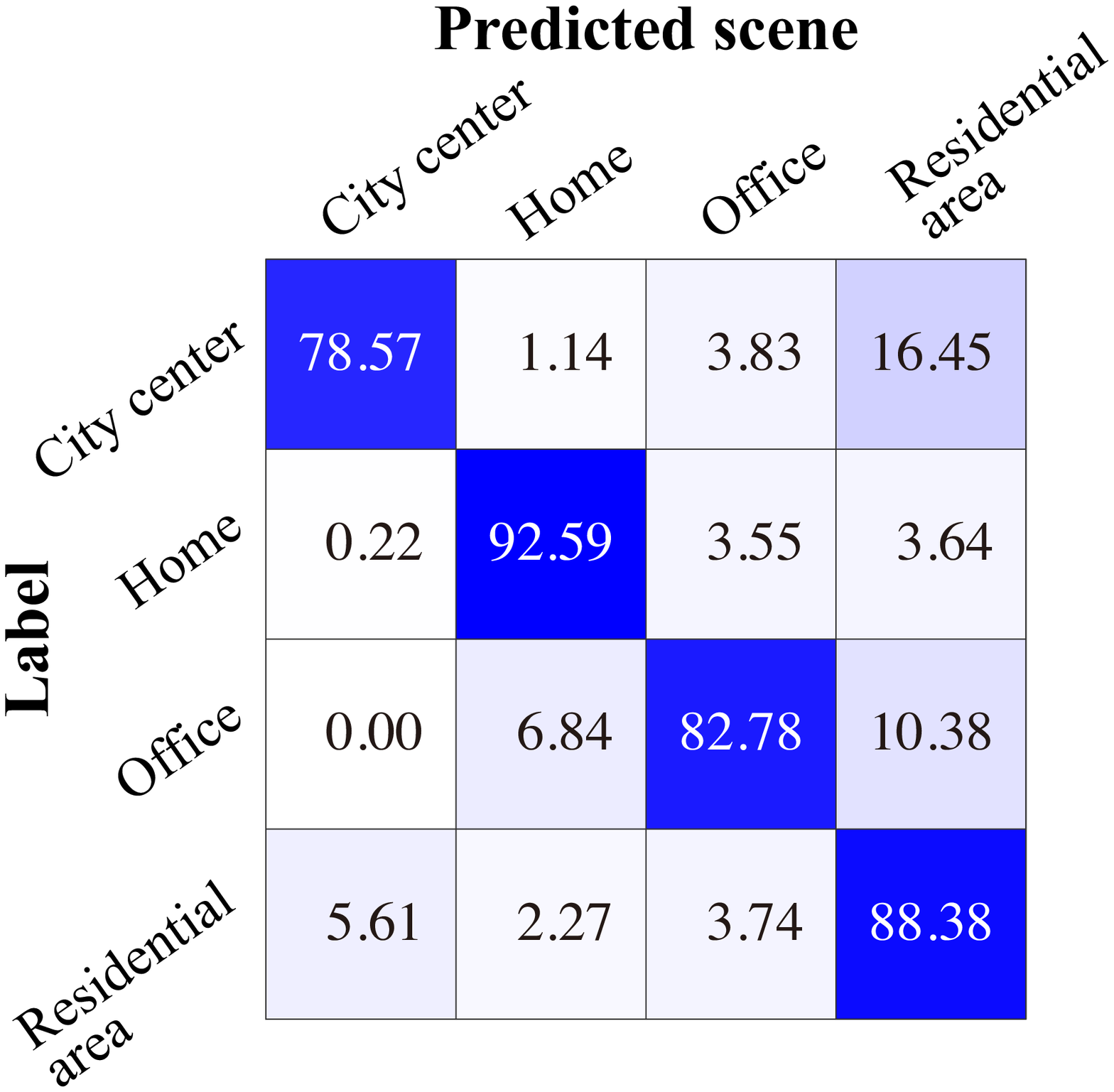}
\vspace{-13pt}
\subcaption{CNN (ASC)}
\end{minipage}
\hspace{5pt}
\begin{minipage}[b]{0.48\linewidth}
\hspace*{-9pt}
\centering
\includegraphics[width=1.11\columnwidth]{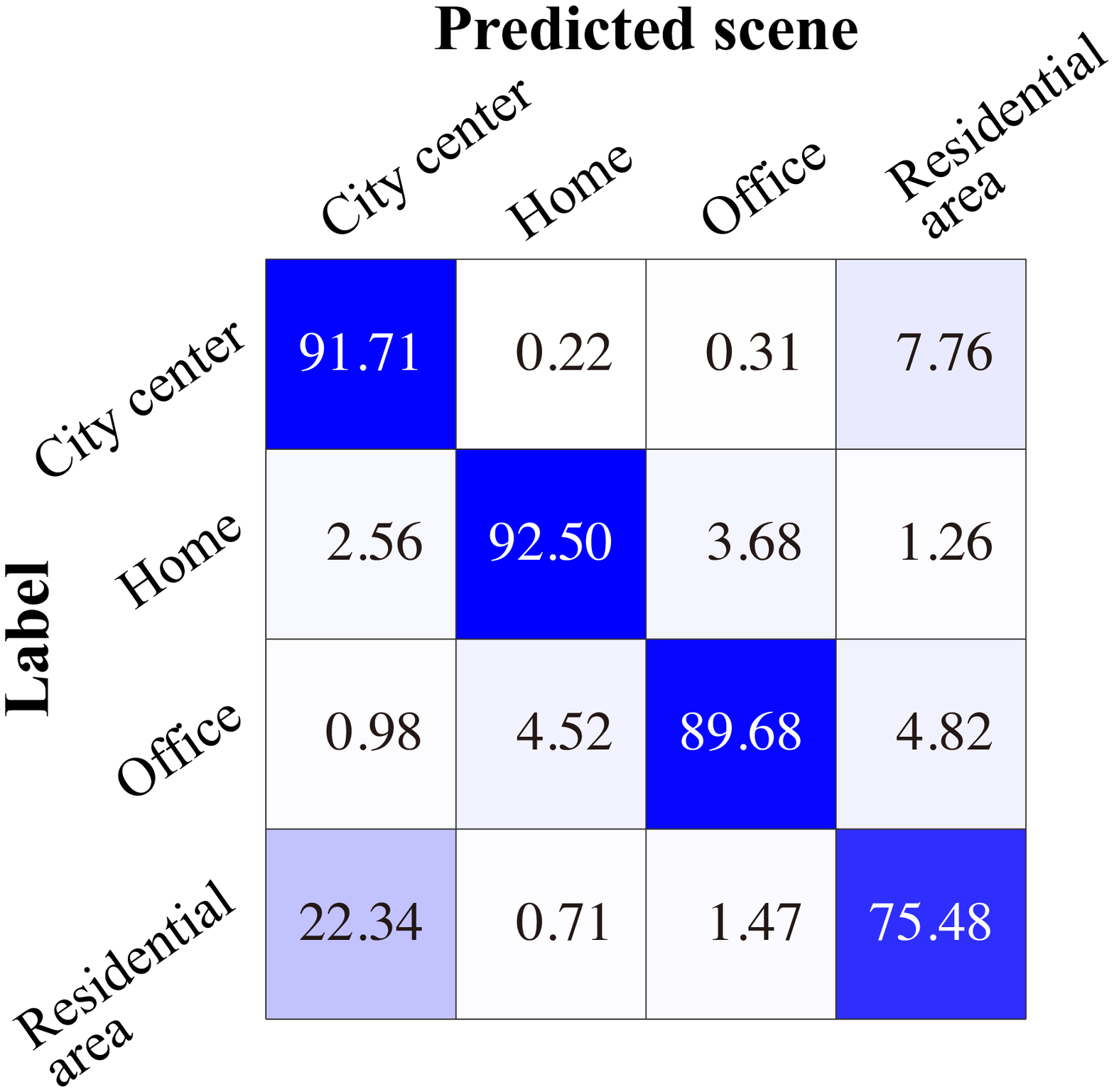}
\vspace{-13pt}
\subcaption{MTL of ASC and SED}
\end{minipage}
\ \\[4pt]
\begin{minipage}[b]{0.48\linewidth}
\hspace*{-9pt}
\centering
\includegraphics[width=1.11\columnwidth]{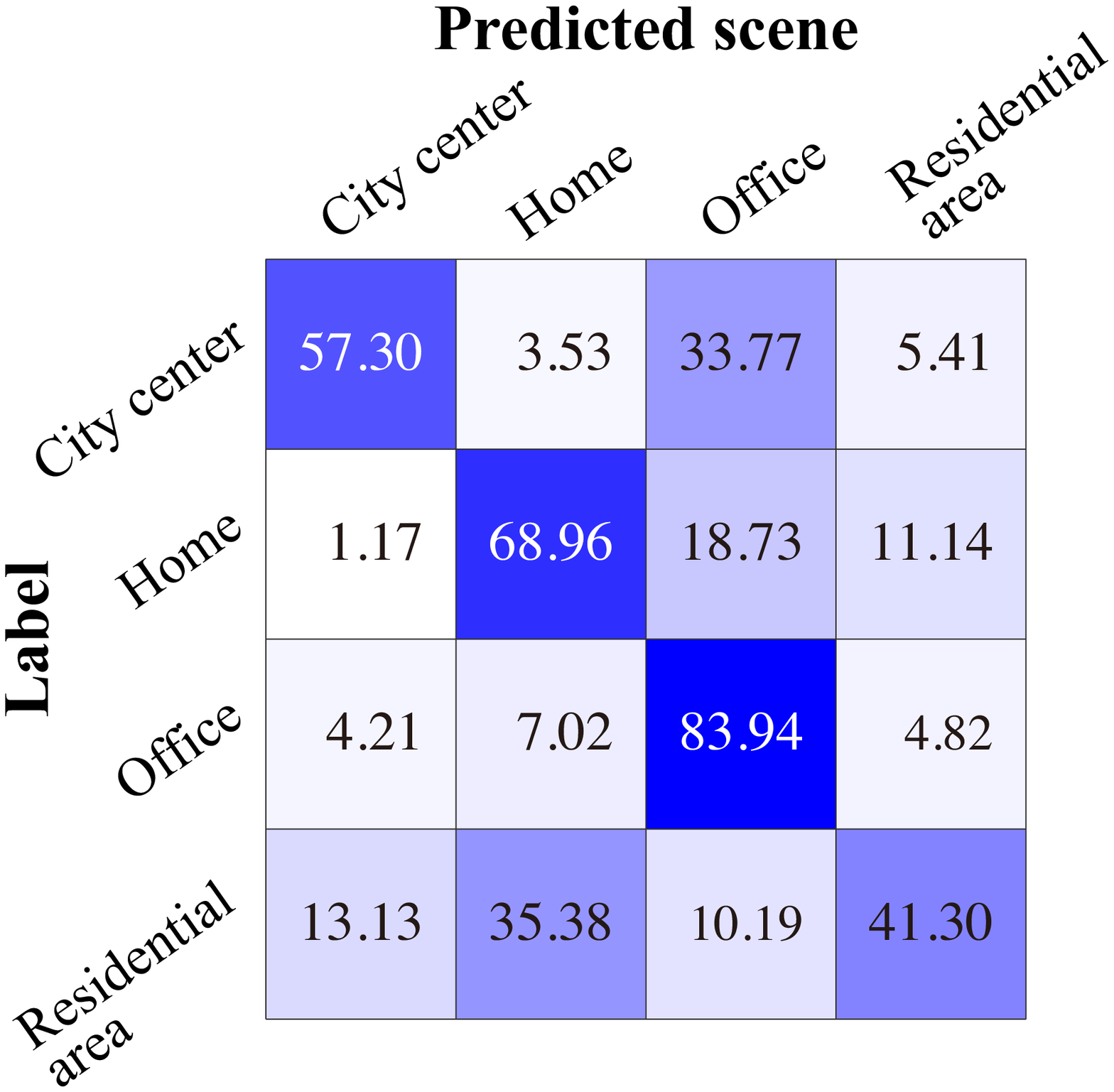}
\vspace{-13pt}
\subcaption{GRL (E1)}
\end{minipage}
\hspace{5pt}
\begin{minipage}[b]{0.48\linewidth}
\hspace*{-9pt}
\centering
\includegraphics[width=1.11\columnwidth]{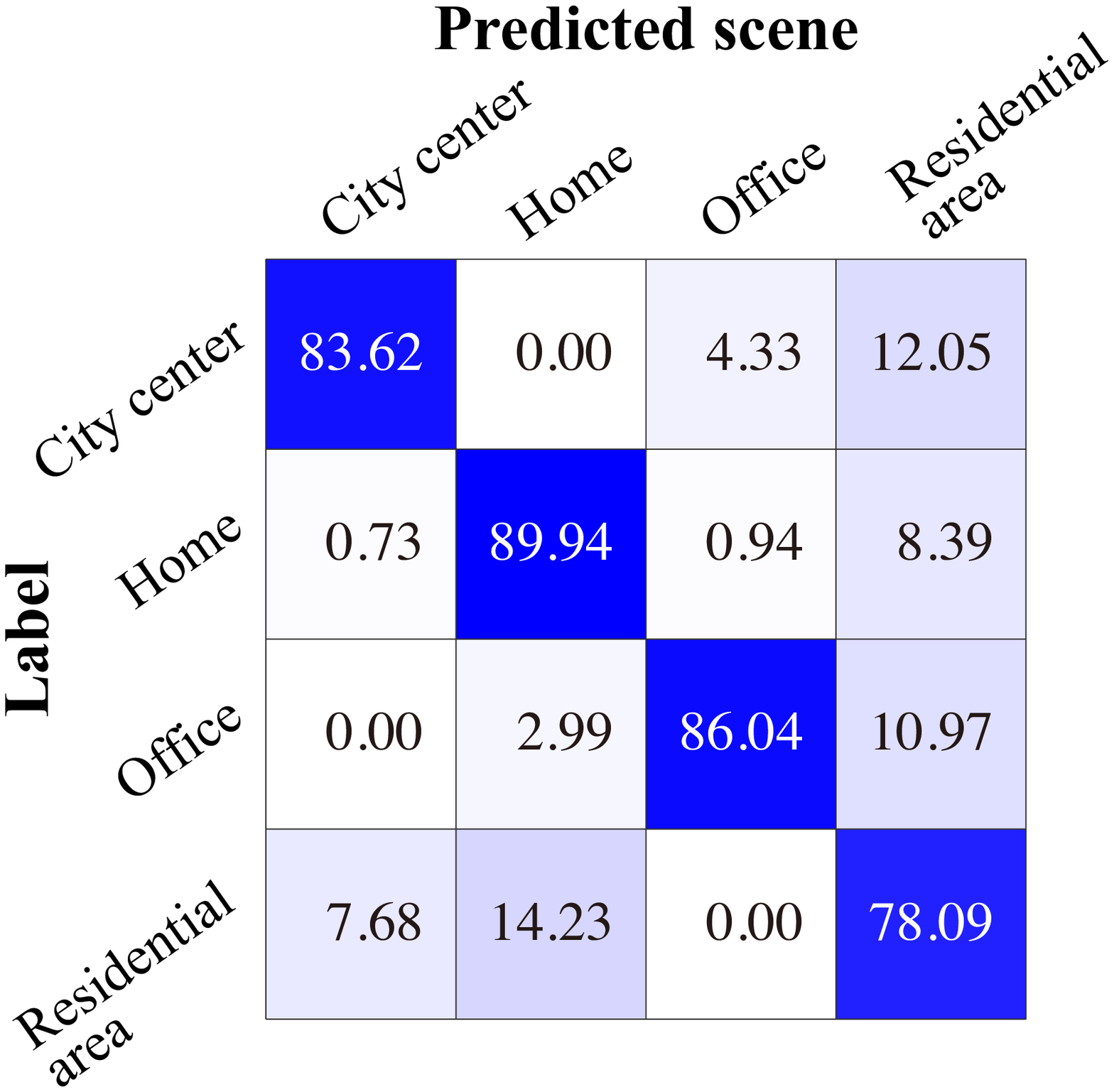}
\vspace{-13pt}
\subcaption{Fake event label}
\end{minipage}
\caption{Confusion matrix for each method in terms of recall (\%)}
\label{fig:confmat}
\vspace{10pt}
\end{figure}
%
%
%
\subsubsection{Details of Performance of SED and ASC}
\label{sssec:detailedresults}
To investigate the event detection performance in detail, we show the Fscores for selected sound events in Table~\ref{tab:eventresult1}.
Note that a similar trend is observed for sound events not shown in Table~\ref{tab:eventresult1}.
This result shows that the Fscores decrease when we train the models with GRL (S1), GRL (S2), and fake scene labels.
This result also supports that information on acoustic scenes is effectively used to detect sound events.
Moreover, the experimental result shows that CNN-BiGRU, which is the single-task-based method, outperforms GRL (S1), GRL (S2), and fake scene labels.
Thus, when detecting most sound events using the single-task-based method, information on acoustic scenes may also benefit SED implicitly.

The detailed classification results of acoustic scenes are shown in Fig.~\ref{fig:confmat}.
Compared with CNN (ASC), the MTL of ASC and SED has fewer misclassifications of \textit{office}.
Moreover, the MTL of ASC and SED still confuses \textit{city center} and \textit{residential area}.
This is because in \textit{city center} and \textit{residential area}, similar sound events tend to occur, as shown in Fig.~\ref{fig:numofinstance}; thus, providing information on sound events yields fewer benefits in distinguishing between these acoustic scenes.
The comparison of CNN (ASC) and GRL (E2) also shows that information on annotated sound events is used in the single-task-based ASC method implicitly.
On the other hand, the impact of unlabeled sound events, such as background noise, on ASC must be further investigated in future work.
%
%
%
\section{Conclusions}
\label{sec:conclusion}
In this study, we investigated how information on acoustic scenes and sound events benefits SED and ASC in detail.
To evaluate this, we applied domain adversarial training based on GRL and fake labels of acoustic scenes and sound events. 
The evaluation experiments conducted using parts of the TUT Acoustic Scenes 2016/2017 and TUT Sound Events 2016/2017 datasets indicated that pieces of information on acoustic scenes and sound events were effectively utilized for SED and ASC, respectively.
Moreover, by comparing GRL- and fake-label-based methods with single-task-based methods, we found that single-task-based methods achieve better performance.
This implies that even when using single-task-based ASC and SED methods, information on acoustic scenes may be implicitly utilized for SED and vice versa.
%
%
\vspace{3pt}
\section{Acknowledgment}
\vspace{7pt}
This work was supported by JSPS KAKENHI Grant Number JP20H00613.
\bibliographystyle{IEEEtran}
\bibliography{IEEEabrv,KeisukeImoto12,INTERSPEECH2022refs}

\end{document}